\documentclass[12pt,a4paper]{article}
\usepackage[cp866]{inputenc}
\usepackage[T2A]{fontenc}

\begin{document}
\def\r{g/cm${}^2$}
\def\d{\displaystyle}

\centerline {\bf Properties of Quasar-Galaxy Associations }

\centerline {\bf  and Gravitational Mesolensing by Halo Objects}

\centerline {Yu.L.Bukhmastova}
\vskip0.5cm
\centerline {\it St.Petersburg State University, St.Petersburg, Russia}
\centerline{bukh\_julia@mail.ru, bukh@astro.spbu.ru   }		 
\vskip 0.4cm
   
{\bf 1. INTRODUCTION}
\vskip 0.3 cm
The problem of physical associatios between distant quasars and nearby 
galaxies has been discussed in the literature for more than thirty years.
Burbidge {\it et al.} [1] published a catalog of galaxy-quasar associations
containing 577 quasars and 500 galaxies. They indicated that quasars show a 
tendency to be located near the halos of normal galaxies mush more often
than expected in the case of chance projections, and that this phisical 
relationship requires an explanation. The fisical connection between quasars
and galaxies was first interpreted as a result of gravitational lensing by
halo stars in the galaxies (microlensing) by Canizares [2]. However, in
subsequent studies [3-5], it was shown that gravitational microlensing cannot 
explain the observations, due to the extremely low surface density of weak
quasars, which should be amplified by microlensing. In addition, Arp [6]
presented several arguments against the gravitational microlensing hypothesis.

(1) Quasars in pairs with nearby galaxies are predominantly separated from 
the galaxy by several galaxy diameters, so that microlenses should occupy a
huge volume of space around the galaxies. The observed number of associations 
requires anomalously large halo masses.

(2) In groups of galaxies, quasar-galaxy associations are more often encountered
for companion galaxies than for dominant galaxies in the group.

(3) Analysis of archival data on the variability of quasars in associations [7]
showed a lack of variability on timescales of several dozen years, in contradiction 
with expectations of the microlensing hypothesis.

As a result, Arp [6] concluded that the redshifts of quasars could have a non-
cosmologcal nature.

Baryshev and Ezova(Bukhmastova) [8] attempted to explain the appearance of quasar--
 galaxy pairs by theorizing that distant active galactic nuclei experienced  
mesolensing by globular clusters in the halos of more nearby galaxies. In this case,
the quasar corresponds to a distant active galactic nucleus whose brightness is amplified
 by several magnitudes. Estimates of the expected number of associations based
on calculatios of the probability that active galactic nuclei will be lensed lead to 
values of $10^2- 10^4,$  assuming brihtness enhancements of $5^m-8^m.$ If the mean 
brightness enhancement is lowered $3^m$, the number of expected associations
grows to $10^5$ (based on the number of active galactic nuclei being $\sim 10^4\hbox {/deg}^2$
for galaxies with apparent magnitudes to $29^m$). In light of these hypotheses,
the arguments of Arp [6] lose their force.

The current paper is a logical continuation of [8]. We do not yet consider the
problem of the luminosity function of quasars [4,5], which will be the focus of the
next paper in this series. Here, we compose a new, appreciable expanded, catalog of pairs.
The number of pairs selected using the new observational material (8382 pairs)
is in agreement with the predictions of theoretical computations. The search
for pairs was carried out in such a way that the linear distance between the galaxy 
and projected quasar does not exceed 150 kpc. We used four criteria for the pair search:

(1) there are the spatial coordinates $\alpha, \delta  \hbox {and } z$ of the quasars and galaxies,
so that the quasar magnitude $m_Q$ is known;

(2) the quasar must be located farther than the galaxy (i.e., $z_Q>z_G$);

(3) the quasar should be projected onto the halo of the galaxy;

(4) the galaxy should have $z_G> 4\cdot 10^{-4}.$

The criteria for selecting new pairs were based on the main assumptions of the gravitational
lensing hypothesis. In particular, this theory led to criteria (2) and (3). 
Criterion (4) was derived in [8].

The structure of the paper is as follows. Section 2 describes the algorithm for 
selecting the pairs. Section 3 is dedicated to a general analysis of the resulting pairs.
In  section 4, presented histograms reflecting an important property of the
associations and attempt to analyze the result obtained. Section 5 is concerned
 with the Hubble diagram for galaxies and quasars, and Section 6 discusses possible 
tests for the adopted model. We briefly summarize our main conlusions in Section 7.

\vskip 0.7cm

{\bf 2. PAIR SELECTION ALGORITHM}
\vskip 0.4cm
Since the quasars are located at greater distances than the galaxies, we will
initially project the quasar onto a celestial sphere with a radius correspoding to the
distance to the galaxy. We denote the smallest distance between the quasar and galaxy
projectios $l=GQ$ (Fig.1). We now solve for the spherical triangle  PGQ with sides
$PG=90-\delta_G, PQ=90-\delta_Q,$ and  $GQ=l.$
According to the cosine theorem,
$$ \cos l=\cos\delta_G \cos\delta_Q \cos(\alpha_Q-\alpha_G)+\sin\delta_G \sin\delta_Q.$$
The linear distance $x$ in kpc is
$ x=z_G{\d\frac{\sqrt{1-\cos^2l}}{\cos l}}5\cdot 10^6$ for $H=60 \hbox{km s}^{-1}\hbox{Mpc}^{-1}.$

Thus, if $x\leq 150\hbox{kpc}$ and $z_G>4\cdot 10^{-4},$ the quasar-- galaxy pair is specified.

The input quasar catalog [9] is available at the web address:

ftp://cdsarc.u-strasbg.fr/cats/VII/207

The galaxy catalog [10] is available at the address:

 http://www-obs.univ-Lyon1.fr

Our application of the above simple selection operation resulted in the new
catalog of associations, which can be found at the web address:

 http://www.astro.spbu.ru/staff/Baryshev/gl\_dm.htm

\vskip 0.7cm
{\bf 3. MAIN PROPERTIES OF THE CLOSE 

QUASAR-GALAXY PAIRS}
\vskip 0.5cm
The selection criteria indicated above are satisfied by 77843 galaxies and 11358 quasars.
Of these, 1054 galaxies and 3164 quasars were included in pairs, making up
a total of 8382 quasar-galaxy pairs. Cases when one quasar is near several galaxies are 
encountered fairly frequently. Examples of such situations are presented in Table 1.
\vskip 0.5 cm

{\scriptsize\begin{tabular}{|c|c|c|c|c|c|c|c|c|}
\hline
1\hfil &2 \hfil&3 \hfil &4\hfil &5 \hfil &6 \hfil&7 \hfil&8 \hfil&9 \\
\hline
\hline
7627& TEX 1322+479& 2.26 &20.5 &PGC0045939 &0.0005 &14.2 &112.6 &0.0002\\
7628& TEX 1322+479& 2.26 &20.5 &PGC0046039 &0.0006 &11.7 &140.7 &0.0003\\
7629& TEX 1322+479& 2.26 &20.5 &PGC0046127 &0.0009 &13.5 &115.6 &0.0004\\
7630& TEX 1322+479& 2.26 &20.5 &PGC0047270 &0.0015 &13.9 &139.1 &0.0007\\
7631& TEX 1322+479& 2.26 &20.5 &PGC0047404 &0.0015 &8.3  &141.3 &0.0007\\
7632& TEX 1322+479& 2.26 &20.5 &PGC0047413 &0.0016 &9.9  &141.9 &0.0007\\
7880& SBS 1400+541& 0.646&17.5 &PGC0049448 &0.0005 &12.4  &46.7  &0.0007\\
7881& SBS 1400+541& 0.646&17.5 &LEDA0140246&0.0006 &-  &0.7   &0.0009\\
7882& SBS 1400+541& 0.646&17.5&PGC0050063  &0.0008 &7.9  &31.1  &0.0012\\
7883& SBS 1400+541& 0.646&17.5&PGC0050216&     0.0009& 10.9&37.5&0.0014\\
7884& SBS 1400+541& 0.646&17.5&PGC0050262&     0.0010& 13.9&63.2&0.0016\\
7885& SBS 1400+541& 0.646&17.5&LEDA0165626&    0.0005& - &15.5&0.0008\\
7886& SBS 1400+541& 0.646&17.5&LEDA0165627&    0.0010& -&38.0&0.0015\\
7887& SBS 1400+541& 0.646&17.5&LEDA0165629&    0.0010& -&48.7&0.0015\\
8324& Q 2315-4230&  2.83 &20.0&LEDA0123614&    0.0538& 16.1&   71.8&   0.019\\
8325& Q 2315-4230&  2.83 &20.0&PGC0071001&     0.0052& 10.6&   64.9&   0.0019\\
8326& Q 2315-4230&  2.83 &20.0&PGC0071031&     0.0053& 11.1&   57.2&   0.0019\\
8327& Q 2315-4230&  2.83 &20.0&PGC0071066&     0.0056& 10.9&   99.8&   0.0020\\

\hline
\end{tabular}}
\vskip 0.5cm
{\small
 {\bf Table 1.} Examples from the catalog of quasar-galaxy associations showing 
several galaxies near a single quasar. The columns contain: (1) pair number in the original catalog,
(2) quasar name, (3) quasar redshift, (4) quasar apparent magnitude, (5) galaxy name, 
(6) galaxy redshift, (7) galaxy apparent magnitude, (8) distance between the galaxy and projection 
of the quasar onto the plane of the galaxy in kpc, (9) $a=z_G/z_Q.$ The table presents data for three
quasars and eighteen galaxies with which they are associated.}
\newpage

{\scriptsize\begin{tabular}{|c|c|c|c|c|c|c|c|c|}
\hline
1\hfil &2 \hfil&3 \hfil &4\hfil &5 \hfil &6 \hfil&7 \hfil&8 \hfil&9 \\
\hline
\hline
895&    Q 0054+0200&       1.872& 18.4&   PGC0003844&     0.0008& 9.6&    137.6&  0.0004\\
934&    Q 0054+0236&       1.654& 17.6&   PGC0003844&     0.0008& 9.6&    139.1&  0.0005\\
948&    Q 0055+0141&       2.232& 18.6&   PGC0003844&     0.0008& 9.6&    123.1&  0.0003\\
983&    Q 0055+0225&       0.373& 18.6&   PGC0003844&     0.0008& 9.6&    119.6&  0.0021\\
995&    UM 294     &       1.914& 17.1&   PGC0003844&     0.0008& 9.6&    146.4&  0.0004\\
1059&   PC 0056+0125&      3.154& 18.6&   PGC0003844&     0.0008& 9.6&    98.8&   0.0002\\
1060&   Q 0056+0009&       0.613& 18.0&   PGC0003844&     0.0008& 9.6&    148.9&  0.0013\\
1066&   Q 0056+0118&       1.1  & 18.1&   PGC0003844&     0.0008& 9.6&    98.8&   0.0007\\
1095&   Q 0057+0000&       0.776& 17.2&   PGC0003844&     0.0008& 9.6&    149.4&  0.0010\\
1106&   Q 0057+0230&       0.716& 18.3&   PGC0003844&     0.0008& 9.6&    90.5&   0.0011\\
1147&   PHL  938   &       1.959& 17.16&  PGC0003844&     0.0008& 9.6&    67.4&   0.0004\\
1149&   Q 0058+0215&       2.868& 18.91&  PGC0003844&     0.0008& 9.6&    71.2&   0.0003\\
1171&   Q 0058+0218&       0.929& 17.7&   PGC0003844&     0.0008& 9.6&    67.0&   0.0008\\
1182&   Q 0058+0121&       1.432& 17.6&   PGC0003844&     0.0008& 9.6&    66.8&   0.0005\\
1193&   Q 0058+0205&       0.6  & 18.5&   PGC0003844&     0.0008& 9.6&    58.3&   0.0013\\
1195&   Q 0059+0147&       1.143& 18.0&   PGC0003844&     0.0008& 9.6&    55.7&   0.0007\\
1255&   Q 0059+0035&       2.545& 18.5&   PGC0003844&     0.0008& 9.6&    95.1&   0.0003\\
1311&   Q 0100+0228&       1.543& 18.4&   PGC0003844&     0.0008& 9.6&    50.8&   0.0005\\
1314&   Q 0100+0146&       1.909& 18.5&   PGC0003844&     0.0008& 9.6&    30.3&   0.0004\\
1315&   UM 301     &       0.393& 16.39&  PGC0003844&     0.0008& 9.6&    32.0&   0.0020\\
1330&   Q 0100+0106&       1.405& 18.5&   PGC0003844&     0.0008& 9.6&    57.3&   0.0005\\
1335&   Q 0101+0024&       1.436& 18.9&   PGC0003844&     0.0008& 9.6&    100.6&  0.0005\\
1344&   Q 0101+0009&       0.394& 17.4&   PGC0003844&     0.0008& 9.6&    116.6&  0.0020\\
1370&   Q 0101+0118&       1.133& 18.61&  PGC0003844&     0.0008& 9.6&    38.3&   0.0007\\
1403&   Q 0102+0036&       0.649& 18.4&   PGC0003844&     0.0008& 9.6&    84.3&   0.0012\\
1408&   Q 0102+0241&       1.509& 18.6&   PGC0003844&     0.0008& 9.6&    55.3&   0.0005\\
1414&   LMA 15     &       2.7  & 0.0 &   PGC0003844&     0.0008& 9.6&    6.7&    0.0003\\
1444&   Q 0103+0234&       1.7  & 18.0&   PGC0003844&     0.0008& 9.6&    49.8&   0.0005\\
1448&   Q 0103+0024&       1.075& 17.4&   PGC0003844&     0.0008& 9.6&    99.7&   0.0007\\
1454&   PC 0103+0123&      3.066& 19.7&   PGC0003844&     0.0008& 9.6&    36.4&   0.0003\\
1467&   Q 0103+0123&       0.782& 18.8&   PGC0003844&     0.0008& 9.6&    37.9&   0.0010\\
1475&   BRI 0103+0032&     4.437& 18.6&   PGC0003844&     0.0008& 9.6&    92.5&   0.0002\\
1486&   Q 0103-0014&       1.629& 18.5&   PGC0003844&     0.0008& 9.6&    144.7&  0.0005\\
1487&   Q 0104+0001&       0.91 & 18.3&   PGC0003844&     0.0008& 9.6&    127.6&  0.0008\\
1493&   Q 0104+0030&       1.874& 18.5&   PGC0003844&     0.0008& 9.6&    96.9&   0.0004\\
1499&   PC 0104+0215&      4.171& 19.73&  PGC0003844&     0.0008& 9.6&    41.4&   0.0002\\
1578&   PKS 0106+01&       2.107& 18.39&  PGC0003844&     0.0008& 9.6&    73.0&   0.0004\\
1606&   Q 0107+0051&       0.966& 19.0&   PGC0003844&     0.0008& 9.6&    106.4&  0.0008\\
1613&   Q 0107+0022&       1.968& 18.3&   PGC0003844&     0.0008& 9.6&    131.3&  0.0004\\
1640&   MS 01080+0139&     0.713& 17.49&  PGC0003844&     0.0008& 9.6&    96.3&   0.0011\\
1643&   Q 0108+0028&       2.005& 18.25&  PGC0003844&     0.0008& 9.6&    134.2&  0.0004\\
1651&   Q 0108+0030&       0.428& 19.0&   PGC0003844&     0.0008& 9.6&    135.8&  0.0018\\
1678&   MS 01094+0242&     0.262& 18.1&   PGC0003844&     0.0008& 9.6&    132.0&  0.0029\\
1682&   UM  87      &      2.343& 17.3&   PGC0003844&     0.0008& 9.6&    126.5&  0.0003\\
1705&   PB 6325    &       0.774& 17.8&   PGC0003844&     0.0008& 9.6&    146.0&  0.0010\\
1708&   PB 6327    &       1.509& 18.1&   PGC0003844&     0.0008& 9.6&    149.5&  0.0005\\
\hline
\end{tabular}}
\vskip 0.5cm
{\small {\bf Table 2.} Examples from the catalog of quasar-galaxy associations showing a number of quasars
 near a single galaxy. The columns contain: (1) pair number in the original catalog,
(2) quasar name, (3) quasar redshift, (4) quasar apparent magnitude, (5) galaxy name, 
(6) galaxy redshift, (7) galaxy apparent magnitude, (8) distance between the galaxy and projection 
of the quasar onto the plane of the galaxy in kpc, (9) $a=z_G/z_Q.$ The table presents data for  the
galaxy PGC 0003844 and 47 quasars with which it is associated.}
\newpage

 One simple explanation for the data can be obtained if the distribution of
galaxies along the line of sight has a fractal nature; i.e., if, in accordance with [11],
we adopt the following parametric representation for the concentration of objects along 
the line of sight between the source and observer:

 $$n_l(R)=0.5n_{ol}\Bigl[\Bigl(\frac{R}{R_0}\Bigr)^{D_F-3}+
\Bigl(\frac {R_s-R}{R_0}\Bigr)^ {D_F-3}\Bigr],$$
where
$n_{ol}(R_0)$ is the concentration of objects at the distance $R_0,$
$R$ is the distance to the lens, $R_s$ is the distance to the source, and
$D_F$ is the fractal dimension of the galaxy distribution.
Usually, a uniform distribution of lenses is obtained if  $D_F=3,$
in which case $n_l(R)=\hbox{const}.$ The observed value for the fractal dimension
is close to $D_F\sim 2$ [12,13].
The character of this relation is such that, if some quasar (according to the Barnotti--
Tyson hypothesis, the nucleus of an active galaxy) is projected onto the halo of a nearby
 galaxy, the probability that the quasar is also projected onto other galaxies 
is enhanced, since the galaxy is part of a fractal structure. The inverse is also
true, since two objects of a single class along the line of sight in a fractal structure 
have equal validity relative to the path of a light ray. This means that cases when ther are
several quasars near a single galaxy should be rare. Examples of such cases are presented
in Table 2. The galaxies PGC 0003290, PGC 0003589, PGC 0001014, PGC 0003238,
PGC 0003721, PGC 0002789, and a number of other are surrounded by tens or hundreds of
quasars.

Why should a distant quasar be projected onto the halo of a more nearby galaxy?
There exist at least three possible explanations.

(1) If the galaxy is very close to the observer, it covers a relatively large fraction of sky.
There is some probability for the chance projection of quasar onto the halo of the nearby galaxy.

(2) If the spatial distribution of galaxies is fractal and at least some fraction of quasars 
are the nuclei of active galaxies (i.e., they belong to the same general class of objects-- galaxies),
there should be a significant number of pairs due to the properties of fractal distribution,
as noted above.

(3) Due to gravitational mesolensing by halo objects with intermediate mass, such as globular clusters,
whose number in galactic halos can reach several thousand [4], some fraction of objects
viewed through these lenses will appear to have anhanced luminosities and will be
interpreted as being quasars [3]. As a result, the clustering of the mass in the halos of nearby 
galaxies ''magnifies'' the light from a distant galaxy that would otherwise 
simply not be visible. The result is the detection of a close quasar-- galaxy pair.
The result is the detection of a close quasar-- galaxy pair. The appearance of 
numerous quasars around a single galaxy can be explained by the large number of globular 
clusters acting as lenses in the halo.

We will analyze the distribution of the galaxies along the line of sight from the observer to the
quasar in order to elucidate the origin of the observed associations.

\vskip 0.7cm

{\bf 4. MUTUAL LOCATIONS OF THE GALAXIES 

AND QUASARS IN ASSOCIATIONS}
\vskip 0.5cm
The distance to the quasar  (which we take to be indicated by its redshift) 
is denoted $z_Q.$  The analogous distance to the galaxy is $z_G.$ We introduce the 
quantity $a=z_G/z_Q,$  which is the normalized distance from the observer to the 
galaxy in a system in which the distance to the quasar is egual to unity. We 
determined $a$ for all the quasar-galaxy pairs. The result is shown as a 
histogram of$x$ as a function of number of pairs with that $a$ value (not to scale).
We obtain four relations for various halo sizes (50 and 150 kpc) and quasar redshifts (Fig.2).

Even without a more careful analysis, we can see that galaxies in associations are 
preferably located either near the observer, $a<0.1,$ or near the quasar, $a>0.9$,
avoiding intermediate distances between the observer and quasar. The appearence
of the first '' tail'' at  $a<0.1$ is not surprising. Most galaxies are located nearby, so that
 the substantial contribution to the number of pairs from such galaxies could
be an observational selection effect. What about the second ''tail'' at  $a>0.9$, 
which is stable to variation of the halo size and quasar redshift?
Similar dependences were obtained for the catalog [1] in [8], based on 241 pairs.
The data have now been expanded appreciably to 8382 pairs, but the dependences have
remained essentially unchanged. Let us try to elucidate their intrinsic properties.
We assume that the following effects are responsible for these dependences to some
degree:

(1) the fractal distribution of galaxies along the line of sight, with a fractal 
dimension close to 2;

(2) gravitational lensing by globular clusters in the galactic halos 
(or by other halo objects displaying clustering with a King mass distribution);

(3) chance projected pairs.

We studied this question using computer simulations. We fixed the galaxy positions
and scattered the quasars randomly, with $z_Q$ from 0.1 to 3. In this way, we
 attempted to remove the effect of possible lensing. There remain to effects that
could lead to the formation of pairs: random projections and nature of the distribution
of the galaxies along the line of sight. Figure 3 shows the result of identifying
pairs  for the case of a random distribution of background quasars. The results 
have changed both qualitatively and quantitatively. The number of pairs has 
dropped dramatically, to 2000 (compared to 8382), and the second tail in the 
distributions has disappeared; i.e., its presence in the observational data is
 not due to random effects.

As a second step in our computer simulations, we randomly specified the positions 
of 77483 galaxies, with $z_G$ from 0 to 0.25, and of 11358 quasars, with $z_Q$
from 0.1 to 3. Figure 4 shows the result of identifying pairs for this case of 
randomly distributed galaxies and quasars. An analysis of this histogram indicates
that about 1200 pairs could be obtained due to random positional coincidences.
However, another qualitative result is more reliable and interesting: in the
case of chance coincidences, the quasars in pairs are not projected onto nearby 
galaxies whose distances are close to those of the quasars; i.e., again, the 
tail at  $a>0.9$ disappears.

Thus, crude estimates of the model relations suggest that about $15\%$ of the
 pairs formed due to random projected positional coincidences. In the framework
of our adopted assumptions, the remaining pairs form as a result of the fractal
distribution of galaxies along the line of sight and due to lensing by halo
objects with a King mass distribution, with the effects of lensing and of the
fractal galaxy distibution complementing each other [8].
 Let us now consider the following selection of 8382 quasar-galaxy pairs. If
some quasar is projected onto the halos of several galaxies, we select the pair 
with the minimum distance between the quasar and galaxy projected onto a single
galaxy, we consider  all these pairs, since several hundred, or even several 
thousand, globular clusters can be located in the galaxy's halo, each of which 
can act as a lens, resulting in several amplified active nuclei of distant
 galaxies, interpreted as quasars. We now construct the same histogram for such 
a sample of 3164 quasars and 1054 galaxies, shown in Fig.5. Qualitatively, the
 results are as befor: there are excesses of pairs with  $ a<0.1 $ and  $a>0.9.$
Note the numerical values for $a>0.9$  ( ''tail'' with  $ a<0.1 $ has a high fraction 
of chance pairs, and is not informative in a numerical sense). It turns out that the
postulated lenses are located primarily in halos less than  50 kpc in size.
The histogram shows that about $83\%$  of the lenses are located in such halos,
whose size does not exceed  50 kpc while the remaining $17\%$ of the lenses are 
located at distances of  50 to 150 kpc. If these lenses are globular clusters, 
they should be located at precisely such distances from the centers of their galaxies,
in accordance with the data of [14].

As an example, we present data for 146 globular clusters in the Milky Way (Fig.6).
We can see that  140 of the  146 globular clusters are located in the halo out to 
distances of 50 kpc, in agreement with the histograms in Figs.2 and 5. The same 
dependences of the number of globular clusters with distance from the galactic
center are characteristic, for example, А754, А1644, А2124, А2147, А2151, А2152
[15], as well as for NGC4874, NGC4889, NGC4472, NGC4486 [16].
\vskip 0.7cm

{\bf 5. THE HUBBLE DIAGRAM $z=f(m)$

 FOR GALAXIES AND QUASARS }
\vskip 0.3cm
Figure 7 presents a Hubble diagram $ z=f(m),$ for the 77483 LEDA galaxies and
11358 quasars. All the quasars are shifted to the left from the straight line 
corresponding to the ''0.2m'' law.

In the context of the gravitational lensing hypothesis, this shift for the
 quasars can be explained as the effect of brightness amplification during lensing. 
If the quasars are the amplified nuclei of active galaxies, they can be shifted
to the strip corresponding to the brightest galaxies by ''removing'' several magnitudes
from their brightnesses. The presence of a distinct lower boundary to the region 
occupied by the quasars can be explained by the fact that there are objects brighter
than $-23^m$ in the initial catalog of quasars.

\vskip 0.4cm

{\bf 6. OBSERVATIONAL TESTS}
 \vskip 0.3cm

Since the observational data for the quasar-galaxy associations are consistent 
with the gravitational lensing model, this raises the question of futher observational
tests  of the model. We propose the following observational tests of the properties
of the quasars that are members of associations.

(1) The expected angular separation of multiple images of quasars due to lensing
by objects such as globular clusters is several milliarcseconds. Therefore, it is
important to study the structure of the compact radio components of quasars in
associations that have sufficient radio  fluxes for VLBI observations. Of course,
the theory does not exclude the possibility of a single image being formed. In
addition, the image splitting could be unobserved due to insufficient dynamical
 range for the observations.

(2) The expected variability of the quasar's brightness due to the motion of the
 globular clusters takes place on time scales of more than a thousand years. 
Variability on timescales of less than a year is also possible, as a consequence 
of microlensing by individual globular cluster stars. 

(3) Since globular clusters are located in the halos of galaxies, absorption lines 
with $z_{abs}$ corresponding to the redshift of the galaxy may arise in the quasar spectra.
It would be interesting to continue the work begun in [17], comparing the properties of 
quasars with absorption lines and quasars in associations. In particular, the well 
known increase in the number of absorption lines with approach toward the quasar could be
associated with the fractal nature of the large-scale distribution of galaxies along the
line of sight. In this case, we should also expect an increase in the number of absorption 
lines with approach toward the observer.

(4) Comparisons of the spectral properties of quasars in associations with those of various 
types of active qalactic nuclei could serve as a probe of the structure of the emission-line
formation region, which could be affacted differently during lensing by King objects.

(5) Analysing the properties of quasar-galaxy associations, we can estimate the fraction of galaxies
of various morphological types in the total number of associations and the distribution of globular
clusters in each type of galaxy.

(6) In the case of quasar with jets, the curvature in the jet trajectories that is not associated with 
real curved motion of the material ejected from the core, but is related to the refraction of rays in 
the presumed lens is possible.

(7) In the hypothesis of gravitational microlensing of quasars, the Hubble diagram presented above 
can be used to estimate the amplification coefficient for quasars, by moving each quasar to the strip
 of brightest galaxies. Knowing the dependence of the amplification coefficient on the position of the 
lens between the observer and source, it is possible to estimate the distance to the postulated lens.

(8) It is possible to devise a computer simulations of the large-scale fractal distribution of galaxies 
and quasars, in order to organize searches for quasar-galaxy pairs and compare the resulting data with the 
catalog of observed associations.
 \vskip 0.7cm

{\bf 7. CONCLUSION}
\vskip 0.3 cm

We have derived a new catalog of close quasar-- galaxy pairs including 1054 galaxies and 3164 quasars 
comprising 8382 pairs. All the galaxies in the associations show a strong tendency to be located close
to either the observer or the quasar, avoiding intermediate positions along the line of sight to
the quasar. This property of the pairs and their considerable number can easily be explained if we
adopt the following assumptions.

1) The quasars in associations could be gravitationally amplified active nuclei of distant galaxies.
There are sufficient numbers of active galactic nuclei to provide the observed number of quasars if there
is a probability of $10^{-4}$ for a brightness amplification by $3^m.$

2) Galaxies on scales $\sim 150$ Mpc  have a fractal distribution with a fractal dimension close to two.
A uniform galaxy distribution makes it difficult to explain the enhanced number of pairs with 
$a>0.9.$

3)  The role of the gravitational lenses can be played by objects such as globular clusters or clusters of dark matter 
characterized by a King density distribution, located primarily in galactic halos at distances of up to 
50 kpc from the galactic center. The importance of the King distribution is that it has a conical caustic that can explain
 the enhanced probability for the galaxies harboring the lenses to be near either the observer or the quasar.
The point and isothermal-sphere lens models that are often considered in the literature give an enhanced probability 
for the galaxy to be located in a central position between the observer and the source, in contradiction with the observed 
properties of the association. The derived histograms indicate that quasars in associations are most often projected onto 
the galactic halo at distances out to 50 kpc, in spite of the fact that the pair selection criteria allowed projections out to 150 kpc.
This provides an additional argument that the relationship between the distant quasars and nearby galaxies
is associated with globular clusters in the galaxy halos.

4) Quasars that were not included in the catalog of assosiations can also be gravitationally 
amplified galactic nuclei, since it is probable that a galaxy surrounded by lensing objects is present 
near the projection of the quasar and is simply not detected. This possibility follows from the above histograms.

The hypothesis that we are dealing with gravitational lensing og the distant nuclei of galaxies can provide a simple physical
 interpretation of the Arp effect; i.e., the observed frequency with which quasar-galaxy associations are 
encountered. The quasar redshifts are cosmological in nature, and do not require any new physics.
\vskip 0.7 cm

{\bf ACKNOWLEDGMENTS}
\vskip 0.3 cm
The author is grateful to Yu.V.Baryshev for constuctive discussions of the problem, and also to D.S.
Bukhmastov for appreciable technical help in preparing the manuscript for publication.

\vskip 0.7 cm
{\bf REFERENCES}
\vskip 0.3 cm
1. G.Burbidge, A.Hewitt, J.V.Narlikar, and P.Das Gupta,

 {\it Astrophys. J.,Suppl.Ser.} {\bf 74}, 675 (1990).

2. C.R.Canizares, {\it Nature} {\bf 291}, 620 (1981).

3. M.Vietri and J.P.Ostriker, {\it Astrophys. J.} {\bf  267}, 488 (1983).

4. E.V.Linder and P.Schneider, {\it Astron. Astrophys.} {\bf  204}, L8 (1988).

5. P.Schneider, J. Ehlers, and E.E.Falko, {\it Gravitational Lenses} 

(Springer-Verlag, New York, 1992).

6. H.Arp, {\it Quasar, Redshifts and Controversies}

 (Cambridge Univ.Press,Cambridge, 1988).

7. W.C.Keel, {\it Astrophys.J.} {\bf 259}, L1 (1982).

8. Yu.V.Baryshev and Yu.L.Ezova, {\it  Astron.Zh. }{\bf 74}, 497 (1997)

[{\it Astron.Rep.} {\bf 41},436(1997)].

9. M.P. Veron-Cetty and P.Veron,{\it A Catalogue of Quasars and Active

 Galactic Nuclei} (European Southern Observatory, Garching,1998),

 ESO Scientific Report Series, Vol. 18.

10. G.Paturel, H.Andermach, L.Bottinelli, et al., 

{\it Astron.Astrophys.,Suppl.Ser.}{\bf 124}, 109 (1997).

11. Yu.V.Baryshev, A.A.Raikov, and A.A.Tron, in {\it Gravitational Lenses 

in the Universe,} 31st Liege International Astrophysics Colloquium, 1993, p.365.

12. Yu.V.Baryshev, F.Sylos-Labini, M.Montuori and L.Pietronero,

{ \it Vistas Astron}{\bf 38}, 419 (1994).

13. F.Sylos-Labini,  M.Montuori and L.Pietronero, {\it Phys. Rep.}

 {\bf 293}, 61 (1998).

14. W.E.Harris, {\it Astron. J.} {\bf 112}, 1487 (1996); 

http://www.physics.mcmaster.ca/Globular.html.

15. J.P.Blakeslee, http://babbage.sissa.it/astro-ph/9906356.

16. J.P.Blakeslee and J.L.Tonry, {\it Astrophys.J.} {\bf 442}, 579 (1995).

17. A.F.Dravskikh, {\it Astron.Zh.} {\ bf 73}, 19 (1996) [{\it Astron.Rep.} {\bf 40}, 13 (1996)].

\end{document}